# Leveraging web resources for keyword assignment to short text documents

Ayush Singhal     Ravindra Kasturi     Jaideep Srivastava     Ankit Sharma

## Abstract

Assigning relevant keywords to documents is very important for efficient retrieval, clustering and management of the documents. Especially with the web cor-pus deluged with digital documents, automation of this task is of prime importance. Keyword assignment is a broad topic of research which refers to tagging of document with keywords, key-phrases or topics. For text documents, the keyword assignment techniques have been developed under two sub-topics: automatic keyword extraction (AKE) and automatic key-phrase abstraction. However, the approaches developed in the literature for full text documents cannot be used to assign keywords to low text content documents like twitter feeds, news clips, product reviews or even short scholarly text. In this work, we point out several practical challenges encountered in tagging such low text content documents. As a solution to these challenges, we show that the proposed approaches which leverage knowledge from several open source web resources enhance the quality of the tags (keywords) assigned to the low text content documents. The performance of the proposed approach is tested on real world corpus consisting of scholarly documents with text content ranging from only the text in the title of the document (5-10 words) to the summary text/abstract (100-150 words). We find that the proposed approach not just improves the accuracy of keyword assignment but offer a computationally efficient solution which can be used in real world applications.

Ayush Singhal, Jaideep Srivastava, Ankit Sharma
University of Minnesota, Twin Cities, USA
E-mail: ayush@cs.umn.edu, srivasta@umn.edu, ankit@cs.umn.edu

Ravindra Kasturi
Amazon.com, Seattle, USA

## 1 Introduction

Assigning keywords (or tags) to documents is extremely important for several practical applications like search engines, indexing of databases of research documents, comparing similarity of documents, document categorization or classification and ontology creation and mapping, to name a few. While this is important for every information retrieval practices in general, the act of classifying or summarizing documents with keywords (tags) has become very important in maintenance of bibliographic databases. As an example, the number of scholarly documents available on the web is estimated to be 114 million English language scholarly documents [Khabsa et al.2014]. Given the size of the database, several academic indexing services such as Microsoft Academic search[1], Pubmed[2] and MNCAT[3] (to name a few) are using keywords to index the documents rather than the full text because of several practical issues[Lamb2008]. The index terms are mostly assigned by experts but in several cases author keywords are also used.

   Although keyword assignment is a well studied problem in the field of information sciences and digital libraries, there are several practical challenges that cannot be addressed with the state of art techniques for automating keyword assignment. The current approaches for keyword extraction leverage the local content informa-tion of the documents and identify keywords from it. These approaches are suitable for the documents with enough text content since the structural, positional and lin-guistic information can be leveraged to determine certain words as keywords for the document. In the literature, these approaches are broadly classified into two sub-categories: (1) keyword extraction and (2) keyword/ key phrase abstraction. Super-vised [Witten et al.1999,Turney2000] and unsupervised approaches [Mihalcea and Tarau2004, Rose et al.2010] for keywords extraction use the local text content of the document.

These approaches, at best, can assign keywords from only those present in the con-tent. Other keywords which provide better description of the document will be com-pletely missed out. While the keyword abstraction approaches mainly deal with key-word selection from a collection of keywords based on the relevance to documents' content. The keywords, therefore, need not necessarily overlap with the text in the document [Browne1996]. Gabrilovich et. al.[Gabrilovich and Markovitch2006] pro-posed an innovative approach for document categorization which uses of Wikipedia knowledge base to overcome the limitation of generating category terms which are not present in the documents. However, this approach uses the entire content of the document and extend the context using Wikipedia. While, the proposed approaches are useful to assign keywords to full text documents, but they are not necessarily applicable to documents with low text content. By low text content, we mean the documents with word length from anything between 5 to 100 words. Scientific pub-lications which contain only a small abstract, documents over the web to which full-text access is withheld, social snippets from media such as Facebook and twitter, short news texts and product reviews on online stores are a few examples of items with low text content. Keyword assignment to such low content document is very important for the purpose of information management as illustrated in the following few examples.

---

[1] http://academic.research.microsoft.com/

[2] http://www.ncbi.nlm.nih.gov/pubmed

[3] https://www.lib.umn.edu/

Table 1: A few examples of document titles which do not try to capture the essence of the document's content.

| Document titles |
| --- |
| Sic transit gloria telae: towards an understanding of the web's decay |
| Visual Encoding with Jittering Eyes |
| BuzzRank ... and the trend is your friend |

Example 1: Keyword assignment for items with 5-10 text words: Sometimes in conference submission portals, the number of manuscript submissions are exceedingly high. In general authors are provided an option to categorize their document at the time of submission. Based on personal experiences of several researchers, cate-gory selection for the manuscript is a confusing process due to conceptual overlaps among several categories. There are several limitations with this approach. Given the overlapping nature of various research topics, category selection for the document can be more appropriately done with the global perspective of various topics instead of the local perspective of the author. Another challenge from the side of the confer-ence organizers is to categorize the documents by reading through the abstract of all the submissions. Categorization would be much more simplified if keywords could be assigned simply by glancing over the title of the documents. This would reduce the computational cost significantly. However, using only document titles may be trou-blesome if the authors provide a catchy title to the document (as shown in table 1). In this work, we demonstrate that this can be accomplished using our proposed ap-proach that leverage information from several open source web resources to assign relevant keywords.

Example 2: Removing non-relevant keywords: One of the main challenges with automatic or even manual keyword assignment to documents is the problem of irrelevant keyword causing drift from the main topic of the document. Often authors may insert keywords which are too specialized for their own work but such keywords may neither be useful for the purpose of categorization or retrieval of the document from a database. In an automated keyword assignment approach, irrelevant keywords may pop-up either due to the nature of the document or the applied algorithm. In either case, it is important to identify and remove keywords which are irrelevant. We propose an automated approach to handle this key challenge.

Example 3: Keyword extraction from short summary text: The problem of keyword extraction from short summary text of a manuscript is important for several practical purposes. One use case is the example of a conference submission system. In this case, it will be useful if keywords can be recommended to authors based on the information from the summary abstract. Another use case is to assign keywords to the permission protected documents on the web. Often such documents do not have full text access. Thus providing keywords, using only the summary abstract information would be very helpful for the authors. Another interesting use case for the summary text (e.g abstract of research articles) based keyword extraction from very long text documents. It is computationally inefficient to assign keyword by scanning the entire text of the document. However, short summary text has its own challenges. In this work, we identify two challenges associated with keyword extraction from short text. Firstly, the text content may not have sufficient occurrences of words to determine their importance in the structure of the short text. Secondly, the short text available for keyword extraction does not have keywords in the text. We solve these problem by incorporating global information about keywords from web resources. So even if the short text does not contain keywords, the proposed approach can generate keywords from the extended text content from the web resources.

The performance of the proposed approaches for the above mentioned challenges, have been evaluated on several real world datasets. We have mainly used scholarly research articles published in peer reviewed computer science conferences. We con-duct extensive experiment to compare the performance of the proposed approach with several relevant baselines. We find that the proposed approach using web resources to enhance the information in the short text of the document significantly improves over the baseline keyword extraction approaches. In particular, we highlight the suc-cess of appending the short summary text with text content from web resources to extract more relevant keywords in comparison to the keywords extracted from the local content of the document. The proposed approach also shows promising results for keyword assignment using only the title of the document.

The rest of the paper is organized as follows. We discuss the background about several open source web resources in Section 2. The related work is discussed in Sec-tion 3. In Section 4 and 5, we discuss the problem formulation and the overview of the work. In Section 6, 7 and 8, we discuss we discuss the various components of the pro-posed approach in detail. Section 9 and 10 comprises of the various experiments and discussion about their results. Conclusions and directions for future research are pro-vided in Section 11. This paper provides a aggregated work of our previous researches [Singhal et al.2013,Singhal2014,Singhal and Srivastava2014a,Singhal and Srivastava, Singhal and Srivastava2014b].

## 2 Related Work

As described earlier, the literature under document annotation can be divided into two broad classes. The first class of approaches study the problem of annotation using ex-traction techniques [Ertoz̈ et al.2003,Frank et al.1999]. The main objective of such techniques is to identify important words or phrases from within the content of the document to summarize the document. This class of problem is studied in the litera-ture by several names such as "topic identification" [Clifton et al.2004],"categorization" [Sebastiani2002,Joachims1998], ''topic finding" [Lawrie et al.2001],"cluster label-ing" [Moura and Rezende2007,Lin1995,Tiun et al.2001,Zamir and Etzioni1998]and as well as "keyword extraction" [Ertoz̈ et al.2003,Frank et al.1999].

Researchers working on these problems have used both supervised and unsupervised machine learning algorithms to extract summary words for documents. Witten et al. [Witten et al.1999] and Turney [Turney2000] are two key works in the area of supervised keyphrase extraction. In the area of unsupervised algorithms for key phrase extraction, Mihalcea and Tarau [Mihalcea and Tarau2004] gave a textRank algorithm which exploits the structure of the text within the document to find key-phrases. Hasan and Ng [Hasan and Ng2010] give an overview of the unsupervised techniques used in the literature.

In the class of key phrase abstraction based approaches. There can be two approaches for document annotation or document classification: single document annotation and multiple document annotation. In the single document summarization, several deep natural language analysis methods are applied. These strategies of docu-ment summarization use ontology knowledge based summarization [Hassan et al.2012, Jain and Pareek2010]. The ontology sources commonly used are wordNet, UMLS. The second approach widely used in single document summarization is feature ap-praisal based summarization. In this approach, static and dynamic features are con-structed from the given document. Features such as sentence location, named entities, semantic similarity are used for finding document similarity.

In the case of multi-document strategies, the techniques incorporate diversity in the summary words by using words from other documents. However, these tech-niques are limited when the relevant set of documents is not available. Gabrilovich et. al. [Gabrilovich and Markovitch2006] proposed an innovative approach for docu-ment categorization which uses of Wikipedia knowledge base to overcome the limi-tation of generating category terms which are not present in the documents. However, this approach uses the entire content of the document and extend the context using Wikipedia.

## 3 Background of open source web resources

In this section we give a brief overview of the different open source web resources used in this work.

### 3.1 WikiCFP

As the name suggests, WikiCFP is a semantic wiki for calls for papers in science and technology fields. There are about 30,000 CFPs on WikiCFP. The knowledge source is used by more than 100,000 world wide researchers every month. WikiCFP is a well organized resource to browse through the CFPs of conference using keyword search. The CFPs are categorized using genealogy information of research topics. In addition to all this, WikiCFP contains the most updated call for papers and expired CFPs are automatically pushed down. Thus the CFP information is timely updated. In this work we have used the "topics of interest" section of the CFPs for mining up-to-date topic information about research.

*3.2 Crowd-sourced knowledge*

Wikipedia[4] is currently the most popular free-content online encyclopedia containing over 4 million English articles since 2001. At present Wikipedia has a base of about 19 million registered users including over 1400 administrators. Wikipedia is written collaboratively by largely anonymous internet volunteers. There are about 77,000 ac-tive contributors working on the articles in Wikipedia. Thus the knowledge presented in the articles over the Wiki are convinced upon by editors of similar interest.

DBpedia[5] is a crowd-sourced community effort to present the information avail-able in Wikipedia in a structured form. This information can be used to answer so-phisticated queries on the Wikipedia database, for instance 'Give me all cities in New Jersey with more than 10,000 inhabitants' or 'Give me all Italian musicians from the 18th century'. The English version of the DBpedia knowledge base currently de-scribes 4.0 million things, out of which 3.22 million are classified in a consistent ontology.

Similar to DBpedia is Yago which in addition to Wikipedia combines the clean taxonomy of WordNet. Currently, YAGO2s [Hoffart et al.2013] has knowledge of more than 10 million entities (like persons, organizations, cities, etc.) and contains more than 120 million facts about these entities. Moreover, YAGO is an ontology that is anchored in time and space as it attaches a temporal dimension and a spacial dimension to many of its facts and entities proving a confirmed accuracy of 95%.

Freebase [6] is another online collection of structured data harvested from sources such as Wikipedia as well as individually contributed data from its users. Its database is structured as a graph model. This means that instead of using tables and keys to define data structures, Freebase defines its data structure as a set of nodes and a set of links that establish relationships between the nodes. Because its data structure is non-hierarchical, Freebase can model much more complex relationships between individual elements than a conventional database, and is open for users to enter new objects and relationships into the underlying graph.

*3.3 Academic search engines*

Academic search engines provide a universal collection of research documents. Search engines such as Google scholar and similar other academic search engines have made the task of finding relevant documents for a topic of interest very fast and efficient. We use the capacity of search engines to find relevant documents for a given query document. We have used the Google scholar search engine[7] and University of Minnesota's MNCAT library search engine[8] for this purpose.

---

[4] www.wikipedia.org/

[5] http://wiki.dbpedia.org/About

[6] https://www.freebase.com/

[7] scholar.google.com/

[8] https://www.lib.umn.edu/

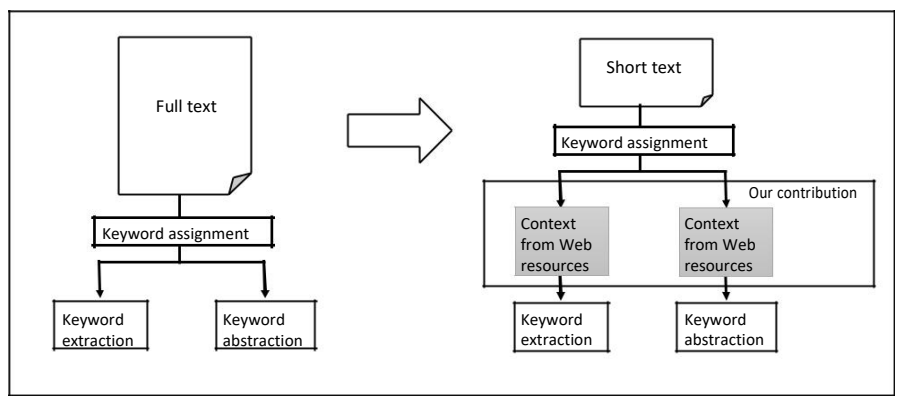

Fig. 1: Figure explaining the big picture of the contributions of our work. Figure (a) is a pictorial summary of the keyword assignment approaches for full text documents, (b) shows proposed model of keyword assignment for short text documents.

## 4 Overview of our work

The unique challenges associated with keyword assignment for short text documents are explained earlier. As a result, computational techniques are required to automati-cally assign relevant keywords to such low content documents.

In this paper, we focus on the problem of automating keyword assignment for low text content documents. In particular, we address the three fold challenges described in section 1. Before moving forward about the details of the proposed approaches, we summarize the flow of the overall framework.

The problem of keyword assignment for low text content documents is addressed in two ways. In figure 1, we compare two different models for keywords assignment problem, one for full text (conventional model) and another for short text (proposed model). In the rest of the paper, we will discuss the model for short text document. We will describe the proposed approaches to leverage content from open source web resources to improve (i) keyword abstraction and (ii) keywords extraction. In the cat-egory of keyword abstraction, we first present an approach that leverage information from web resources like WikiCFP and Wikipedia for automating keyword selection for such document. Generation of keywords from the web resources helps to generate keywords which are up-to-date with the current scientific terminologies. The problem of low text content of the document is overcome by generating additional content us-ing the web resources like academic search engine. The additional content is known as 'global context' because the information is generated using external global re-sources rather than the 'local' text content of the document. The new text content of the document is then used to select keywords from the list of keywords generated us-ing crowd sourced web resources. In the second approach, we fully automate the step of keyword assignment by assigning keywords from crowd sourced topics available in crowd sourced knowledge bases like Dbpedia, Yago and Freebase. The relevance of the keywords to the document is automatically inferred using a web-distance based clustering approach. Non-relevant keywords are detected and removed in an unsuper-vised manner. The proposed approaches are tested on real world dataset from DBLP, the approach, however, is generic enough to be used for various types of documents like news articles, patent documents and other documents where keywords needs to be assigned using only the short text content.

In the category of keyword extraction, we propose a novel model for automatic keyword extraction from the low text content of the document. We show that the quality of the assigned keywords is improved by incorporating relevant context from the web. In addition to the text content of the document, the text content generated from the web adds global information about the document's topic and thus improves keyword extraction. The hypothesis is tested on a real world document corpus. We evaluate the performance of three well known keyword extraction techniques and compared the impact of adding the proposed web-context (described in Section 5) to the local content of the document.

Since, text content generation using open source web resources is used to enhance both the keyword abstraction and extraction, we first describe this step in the next section. For the sake of convenience, we refer to this step as web context extraction.

## 5 Web context extraction

Given the 'short text' $S^0$ (which can be 1-10-word length), the expanded context is generated using open source web resources such as the academic search engine. As shown in figure 5, the context of the 'short text' is expanded using the results obtained by querying the web corpus using an academic search engine. The 'short text' is used as a query to the search engine. For the query $S^0$, the search engine retrieves 'relevant' ranked results. The web context for the input 'short text' is created from the text in the titles of the retrieved documents. There are several other resources in the retrieved results that can be used (like snippets, author information etc) but those are not included in the web context because not all academic search engines offer these features. The final context is created by concatenating the text in the titles of the top-n documents. The value of n is not fixed and can be a parameter to the approach. In the later section, the results are evaluated by varying the value of k. The text is tokenized using space delimiter. Duplicate titles are excluded from the web context. As a basic step in text mining, the tokens are pre-processed by applying stop-word filter, non-alphabetic character removal and length-2 token removal. In the rest of the paper, we will refer to this context for a 'short text' ($S^0$) as web-context (W C($S^0$)) for the sake of convenience and consistency.

## 6 Keyword abstraction

### 6.1 Finding keywords for documents with 5-10 text words

In this section we describe a three step approach taken to automatically assign key-words to a text document given only the title information for the document. The three steps (shown in figure 2) are enumerated in the following subsections.

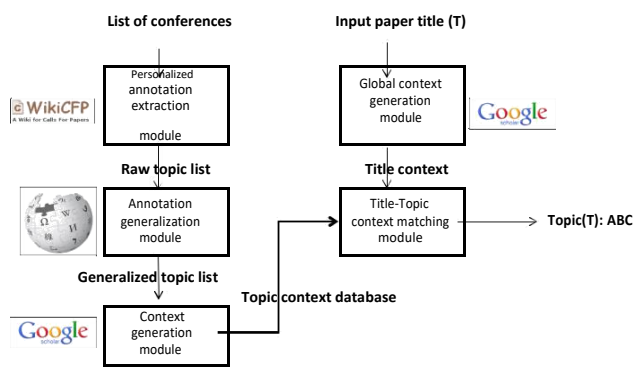

Fig. 2: A systematic framework of the proposed approach

### 6.1.1 Topic Generation

Automatic topic generation is a challenging task. In most of the literature, this step of document annotation involves expert knowledge to generate topics to be assigned to documents. In this work we show a simple and novel approach to automate the task of topic generation using open source media such as WikiCFP and Wikipedia. The proposed topic generation involves two step (as shown in figure 2)

Personalized Annotation retrieval: In this step, we first identify the human created labels for scientific research topics. At the first step, a list W consisting of various venues of research publications (conferences and Journals) is created in an unsupervised manner. In this work, we have narrowed down the domain to data mining research venues. For each venue in this list W, personalized annotations are retrieved from its CFP (Call for papers) from WikiCFP (highlighted with blue box in figure 3. As mentioned earlier, WikiCFP provides a feature for obtaining the CFPs for any research venue. The organized knowledge in this database helps to retrieve various forms of information about the venue such as its categories, deadline, date, CFPs and related venues. We use the last two features to create a database of personalized annotations for research venues in list W . For the venues where the CFP informa-tion is unavailable its related venues are queried in the similar manner. The output of this step is a database D consisting of human generated topics for paper submission. Database creation from WikiCFP assures that the research topics are up-to-date with the research trend. It also enhances the diversity of topics in the database.

Wikifying Personalized Annotations: The result of the previous step is a database D consisting of human generated topics. However, the database contains noise in sev-eral forms such as repetition of research topics, irrelevant topics, highly specialized topics. As an example, "mining of web data" and "web based mining" refer to the same topic "web mining" however they were expressed differently in different CFPs due to personalization by different individuals. Given the complications of natural language interpretation, construction of a refined list of topics from the topics in the database D cannot be done by fusion or intersection of the duplicate topics. In order to remove such noise from the database D, we have used the crowd sourced intelli-gence of Wikipedia to refine and generalize them. This step of using Wikipedia for refining the personalized topics is termed as Wikifying. The following steps are used for Wikifying.

Fig. 3: A snapshot from WikiCFP showing the personalized topic list in the conference CFP.

– Wiki Querying- Each topic in D is queried as a Wikipedia item in the Google search engine. Given a query q in the Google search engine as Wikipedia: q, it returns a list of results from Wikipedia titles with which are related to q.
– Top-N extraction- Next, we extract the top-N Wikipedia results (document titles) for the query q from the Google search page (as shown in the figure 4).

The idea behind wikifying the original annotations from CFPs is to bring in an element of consensus from crowd sourced knowledge of Wikipedia. As an output of this step, we get a refined database $D^0$ which contains refined topics after wikifying the topics in database D. We call these topics as wikified annotations.

### 6.1.2 Using the web context

Using the approach discussed in section 5, we derive the additional text content for each keyword (topic) in the list of keywords ($D^0$) by using the keyword as the input to generate the web context ($WC_{topic}$). Similarly, the additional text content for the input document ($WC_{di}$) is derived using its title as the input 'short text' for web context generation.

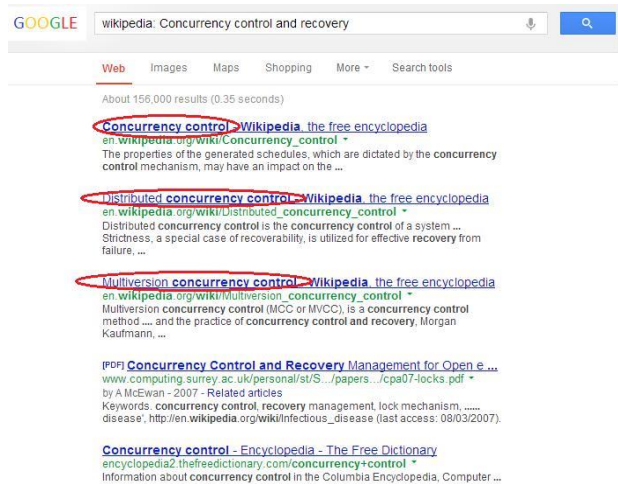

Fig. 4: A snapshot of Google web page showing the Wikification of the input query.

### 6.1.3 Topic Ranking

As stated in the previous step, each topic in the corpus $D^0$ is described by its web context $WC_{topic}$. Similarly, the title of a given document is described by its web context $WC_{di}$. The topics in the corpus $D^0$ are ranked for an input document title $S^0$ using three text similarity computation models namely, TF-IDF [Wu et al.2008], LDA-TFIDF model [Blei et al.2003] and LSI-TFIDF model [Dumais et al.1995].

Ranking: The ranking of the topics in $D^0$ for a given document is obtained by computing the cosine similarity between the features derived from the web context of input document and the keywords in $D^0$. For a given document, the topics $\epsilon$ $D^0$ are ranked in the decreasing order of their cosine similarity scores.

### 6.2 Automating keyword abstraction and de-noising

In this section, we a different approach for keyword abstraction. This approach also includes an approach to de-noise the predicted keyword list. Similar to the previous section, the only text information about the document is the text information in its title.

The task of fully automating keyword assignment is accomplished by selecting keywords from the topics/keywords in the crowd sourced knowledge bases like Db-pedia, Yago and Freebase and finally the non relevant tags are removed a fully un-supervised approach. There are three main components of this approach: (1) Context expansion using academic search engine (similar to web context generation in previ-ous section), (2) candidate keyword generation using crowd-sourced knowledge and (3) de-noising keywords using web-based distance clustering technique.

### 6.2.1 Using the web context

Given the title text of a document (di) as the only text information for the document, its additional text content is generated using the title text as the input for web context extraction (described in section 5). It is referred as web context (W C($S^0$)).

### 6.2.2 Tag generation

In this section, we describe the procedure to utilize crowd-sourced knowledge to generate keywords from the expanded context W C($S^0$). As described earlier, the crowd-sourced knowledge is available in well-structured formats unlike the unstructured web. The structured nature of knowledge from sources such as DBpedia, Freebase, Yago, Cyc provides opportunity to tap in the world knowledge from these sources. The knowledge of these sources is used in the form of concepts and named entity in-formation present in them since the concepts and named entities consists of generic terms useful for tagging. We have used the AlchemyAPI [AlchemyAPI2013] to access these knowledge bases. A tool such as this provide a one-stroke access to as these knowledge bases at once and returns a union of results from all the various sources.

Given the expanded context (W C($S^0$)) as the input to the AlchemyAPI, which matches the W C($S^0$) against the indices of these knowledge sources to match W C($S^0$), using the word frequency distribution, with concepts and named entities stored in the knowledge bases. The output for an API query W C($S^0$) is a list of concepts and named entities. Using the open source knowledge bases and the word frequency in-formation from the input, the API returns a list of concepts related to the content. The named entity list returned for a query W C($S^0$) consist of only those named entities of type 'field terminologies. There are other types of named entities such as 'person name', 'job title', 'institution' and a few other categories but those are not generic enough to be used as keywords. The concepts and named entities for W C($S^0$) together form a keyword cloud T.

Figure 5 highlights a keyword cloud consisting of keywords generated using the above described technique. As shown in the figure, the keywords are weighted based on the word distribution in C($S^0$). This example also shows a few keywords like 'cerebral', 'cortex', 'genomic' that appear to be inconsistent with the overall theme of the cloud for C($S^0$). The next step describes an algorithm to handle such situations in tagging process.

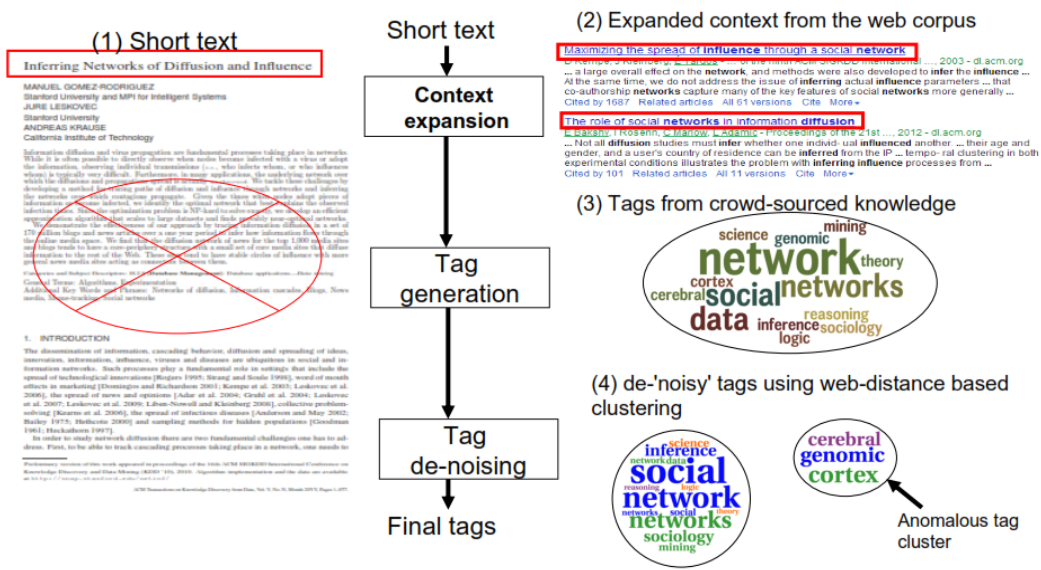

Fig. 5: A systematic framework of the proposed approach. An example is illustrated to explain the proposed approach.

### 6.2.3 Keyword de-noising

As described in the previous step, the keyword cloud T for $C(S^0)$ may contain some inconsistent or 'noise' keywords in it. In this section we describe an algorithm to handle the problem of 'noise' in the keyword cloud. This step is therefore termed as keyword de-noising.

Given the keyword cloud T for $C(S^0)$, noisy keywords are pruned in the following manner. The keywords in T are clustered using a pairwise semantic distance measure. Between any two keywords in T , the semantic distance is computed us-ing the unstructured web in the following way. For any two keywords $t_1$ and $t_2$ in T, dis($t_1$; $t_2$) is defined as the normalized Google distance(NGD)[Cilibrasi and Vitanyi2007]:

$$NGD(t_1,t_2) = \frac{max\{logf(t_1),logf(t_2)\}-logf(t_1,t_2)}{logM-min\{logf(t_1),logf(t_2)\}}$$

where M is the total number of web pages indexed by the search engine; f($t_1$) and f($t_2$) are the number of hits for search terms $t_1$ and $t_2$, respectively; and f($t_1$, $t_2$) is the number of web pages on which both $t_1$ and $t_2$ occur simultaneously.

Using the NGD metric, a pairwise distance matrix(M) is generated for the key-word cloud T. The pairwise matrix M is then used to identify clusters in the keyword cloud. The cloud is then partitioned into two clusters using different hierarchical clus-tering techniques. Here, we assume that there is at least one 'noise' tag in the tag cloud T. Out of the two clusters identified from the keyword cloud T, the one cluster with majority tags is called a normal cluster whereas the other cluster is called as out-lier cluster (or noisy cluster). In case of no clear majority the tie is broken randomly.

The algorithm is illustrated through an example shown in figure 5 step 4. This step shows that the keywords generated in step 3 are partitioned into two clusters as described above. The keywords in one clusters are semantically closer than the key-words in the other clusters. As shown in this example, the outlier keywords 'cerebral', 'cortex', 'genomic' are clustered together while the remaining normal keywords clus-ter together. Since the former is a smaller cluster, it is pruned out from the keyword cloud. Lastly, the final keywords consist of only the keywrods in the larger cluster.

## 7 Keyword extraction from the summary text of a document

In the domain of document-oriented keyword extraction, the recent methods focus on three approaches namely, (1) natural language processing based techniques (2) statis-tical co-occurrence based techniques (3) world knowledge based techniques. In order to familiarize the readers with the above mentioned approaches, we present a brief description about the popular algorithms in each of the above categories. Although these techniques are not the contributions of this work but we will extensively use these techniques in the proposed model for keyword extraction.

### 7.1 TextRank[Mihalcea and Tarau2004]

Under the natural language based techniques, TextRank is the most popular technique for document-oriented keyword extraction. TextRank is based on term selection from the text based on the part-of-speech (POS) tagging of the terms and then applying a series of syntactic filters to identify POS tags that are used to select words to evaluate as keywords [Rose et al.2010]. Using a fixed-size sliding window approach, co-occurring words within the window are is accumulated within a word co-occurrence graph. The candidate keyword are selected from the co-occurrence graph by ranking the words based on a graph based ranking algorithm (TextRank-similar to pageRank [Page et al.1999]). The top-ranking words are selected as key-words. Multi-word keywords are formed by combining adjacent keywords.

### 7.2 Rake [Rose et al.2010]

Under the statistical co-occurrence based techniques, the Rapid automatic keyword extraction (RAKE) is most popular for document oriented keyword extraction. RAKE is an unsupervised, domain-independent and language independent method for extracting keywords from individual documents. Keyword extraction is done by means of ranking the non stop-words (a list of most common English words). The phrases in the non stop-words are identified across the document and then scored based on the proposed metrics. The input parameters for RAKE comprise of the document text and a set of stop-words, a set of phrase delimiters and a set of word delimiters. RAKE uses stop words and phrase delimiters to partition the document in candidate keywords. Based on the co-occurrence of words within the candidate keywords, the word co-occurrences are identified. This approach for identifying co-occurring words save the computation cost of using sliding window technique used in TextRank. Fi-nally, the word association is measured in a manner that adapts to the style and content of the text and therefore important for scoring the candidate keywords.

### 7.3 Alchemy [AlchemyAPI2013]

There are several techniques used to leverage world knowledge to extract keywords. In addition to the local content and style information from the document to ex-tract keywords, these approaches utilize information from crowd sourced corpus like Wikipedia, Freebase, Yago and other similar corpus to identify keywords. The AlchemyAPI[AlchemyAPI2013] uses statistical algorithms and natural language pro-cessing technology, in addition to the world knowledge using various crowd-sourced corpus, to extract keywords from the text of the documents. The scoring of the can-didate keywords is influenced by incorporating statistical information from the world knowledge corpus.

### 7.4 Adding web context

In addition to the local content of the document, the additional information with the author to assign keywords comes from the various other related literature. An author, generally assign keywords that highlight both the local content of the document as well as the relation of the document with the other topics in the category of his/her document. So the author assigned keywords are a balanced mixture of both the local content and the global content. Here, we generate the web context for a document ($d_i$) using its title text as the input for the web context extraction step (described in section 5).

## 8 Experimental analysis for keyword abstraction for documents with 5-10 text words

In this section, we discuss the experiments and results for the keyword abstraction framework for both the approaches described in sections 6.1 and 6.2. We first describe the test dataset, the ground truth, the baseline and the evaluation metric used for evaluation of the proposed approach.

### 8.1 Test dataset description

For the purpose of evaluating our approach, a test set consisting of 50 research doc-uments from top tier computer science conferences was constructed. The 50 papers were selected to capture the variety of documents in computer science research. Sev-eral of the documents had catchy titles (examples given in table 3 and the titles were never intended to convey the core idea of the document. In our algorithm, we used only the title information as the input to the algorithms.

For the proposed approach, the aggregated keywords, from the conferences in which these 50 papers were published, totals to 777 keywords. The keywords for the test documents was selected from these keywords.

### 8.2 Ground truth

In absence of any gold standard annotations for the test documents, the ground truth for the documents was collected from the author assigned keywords to these docu-ments. We collected this information by parsing the 50 documents in the test set. We assume that the keywords assign by the authors are representative of the annotations for the document. The proposed approach and the baselines were evaluated on this ground truth.

### 8.3 Baseline approach

We have compared the performance of the proposed approaches with two baselines which use only the local content information for topic assignment. The first base-line uses only the title and abstract information about the document to identify rel-evant topics from a given list of topics. For the keyword selection algorithm, the keywords are selected from a list of keywords available in faceted DBLP project [DBLP]. This list contains the most popular author keywords assigned to a minimum of 100 research articles in the DBLP dataset [Ley and Reuther2006]. The keywords in the list, unlike those used in the proposed approach, are actually human generated keywords/topics. The second baseline uses the entire content of the documents (title, abstract and all the other section). While taking into account the full content of the documents, we removed the keywords assigned to the paper. We used a tf-idf model to vectorize each of the test document in terms of the keywords. Using the top-10 highest tf-idf values, we assigned 10 keywords to each document from the list of given keywords.

For the fully automated keyword abstraction algorithm, we compare the perfor-mance of the proposed approach with a baseline using the full text content of the test documents. The full text information is generated from the pdf versions of the test documents. The PDF documents were converted to text files using PDF conversion tools. As a basic pre-processing step, stop-words, non-alphabetical characters and special symbols were removed from the text to generate a bag of word representa-tion of the full text. For the purpose of comparison, the full text context was used to generate keywords using the proposed approach and at the final step, de-noising of tags was done using the proposed algorithm. The purpose of this baseline is to see the effectiveness of using the global context as a replacement for the full text content of the document.

Table 2: Table shows comparison of recall for the baselines ($BS_{abs}$; $BS_{whole}$) using only the local context information for document summarization and the proposed approach ($PA_{T\ F\ IDF}$ ; $PA_{LDA}$; $PA_{LSI}$ ) which uses global context generated using search engines in the top-10 results.

| Annotation techniques | $BS_{abs}$ | $BS_{whole}$ | $PA_{TFIDF}$ | $PA_{LDA}$ | $PA_{LSI}$ |
|---|---|---|---|---|---|
| Recall for top-10 | 0.50 | 0.50 | 0.56 | 0.50 | 0.50 |

### 8.4 Evaluation metrics
Given that evaluation by experts is resourceful and time consuming, hence challenging, we evaluate the results of the proposed approach for keyword selection (ref: section 6) using the recall metric.

### 8.4.1 Recall
Recall is used to compute the proportion of the ground truth results that was correctly identified in the retrieved results. Each document is marked to be labeled correctly if atleast one of the retrieved/predicted keywords exactly matches with any one keyword in the ground truth for the document. The proposed approach and the baselines were evaluated on the same metric.

### 8.4.2 Jaccard index
The Jaccard similarity between two sets A and B is defined as the ratio of the size of the intersection of these sets to the size of the union of the sets. It can be mathemati-cally stated as:

$$J(A, B) = \frac{|A \bigcap B|}{|A \bigcup B|}$$

We compare the Jaccard index of the predicted tags (with the ground truth tags) and the baseline tags (with the ground truth tags). The Jaccard index is averaged over the total number of documents in the test dataset.

### 8.4.3 Execution time
The final metric for comparing the proposed approach with the baseline is the execution times. Since the main overhead of the approach is in the first step of tag generation due to difference in sizes of the input context. The execution time is com-puted as the time taken in seconds to generate tags for the 50 test documents given their input context. For the proposed approach the context is derived using web in-telligence whereas for the baseline the context is the full text of the test document. Pre-processing overheads are not taken into account while computing execution tim-ings.

*8.5 Results and discussion for keyword abstraction without de-noising*

This section is sub-categorized into two sections. The first section discusses the quan-titative evaluation of the proposed work. In the next section, we qualitatively discuss our results for a few examples of documents with catchy titles.

*8.5.1 Quantitative evaluation*

In this section, we discuss the results of the experiments performed in the previous section. Table 2 gives a summary of the recall in the top-10 results obtained using the proposed approaches and the baselines. As shown in the table, the recall in the top-10 improves (0.56 vs 0.50) in case of the proposed approach (using TFIDF model for comparing content similarity). It can also be seen that the recall using the proposed approach is comparable to the baseline approaches (approx. 0.50 in both cases). The comparable results signify the effectiveness of the fully automated approach to iden-tify topics for research documents. Thus, given only the title information about a research document, we can automatically assign topics to the document which are very close to human assigned topics.

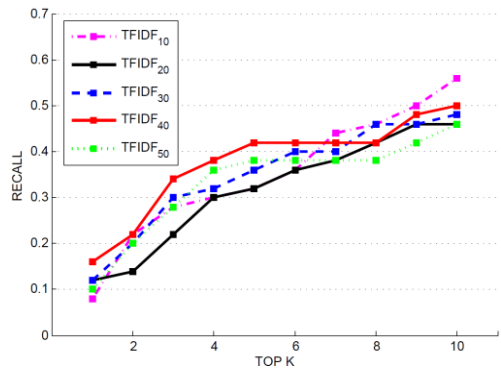

Fig. 6 Recall@k using TFIDF.

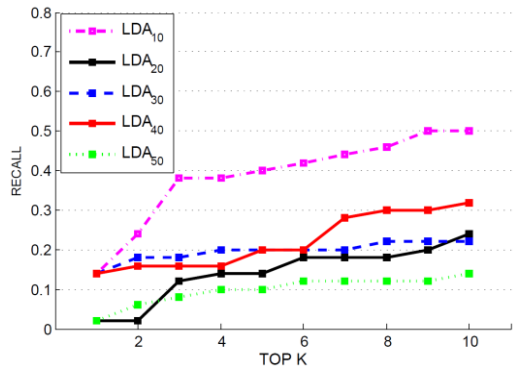

Fig. 7 Recall@k using LDA.

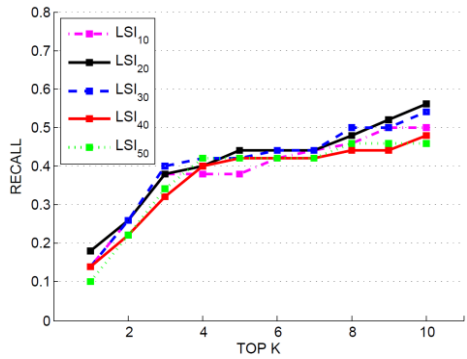

Fig. 8 Recall@k using LSI.

In the next set of experiments, the global context size was varied from 10 through 50 in steps of 10. In the figures 6, 7, 8, we show the variation of recall in the top-k results of the different proposed approaches. In these figures, we also compare the recalls when the global context size is varied. In the figure 6, the recall increases monotonically in all the different global context sizes. The highest recall (0.56) us-ing TFIDF model occurs in the top-10 and using the global context of size 10. The context size of 40 gives the next best recall. However, we also observe that adding extra context (e.g green curve for context size 50-T F IDF$_{50}$), the recall is lower in comparison with other context sizes.

In figure 7, we observe a significant difference in the performances by varying the context size. The highest recall attained using LDA model is 0.50 and it happens when the context size is 10. From this figure, we see that the LDA model is very sensitive to the content information. As shown in the figure, adding extra content (green curve- LDA$_{50}$), the recall is very low ($< 0.20$).

Table 3: Table showing the topics assigned to the documents with catchy titles.

| Document titles | Our approach | Ground truth |
|---|---|---|
| Sic transit gloria telae: towards an understanding of the web's decay | link analysis, adversarial information retrieval, bayesian spam filtering | Link analysis, Web information retrieval, Web decay, dead links |
| Visual Encoding with Jittering Eyes | semantic memory, bag-of-words model, motion analysis, computer vision | Information retrieval, personal information management, human-computer interaction, World Wide Web use |
| BuzzRank … and the trend is your friend | Pagerank, web community, e-social science | Web graph, Web dynamics, PageRank |

In the case of LSI model for similarity evaluation, figure 8 shows a similar trend in performances using different context sizes. As shown, the recall for context sizes of 40 and 50 (shown in red and green curves respectively), is lower in comparison to the recall using context size of 10, 20 or 30. However, the highest recall (0.58) is attained when the context size is 20.

*8.5.2 Qualitative evaluation*

In this section we discuss the results of the proposed approach by qualitatively ana-lyzing the results of the proposed algorithm. Using a quantitative measure like recall fails to account for the subjective accuracy of other topics assigned to a document in our top-10 results when compared to the ground truth by excluding the exact match scenario. Here we analyze the results in a subjective manner.

Table 3 shows the results of summarization using the proposed approach and the ground truth keywords. This is special category of documents where the titles of the document are "catchy" and do not intend to display the core idea of the document. Our algorithm uses only this title information to generate the context and find topics suitable for these documents. The column two shows some of our results in the top 10. For the document titled as "Sic transit gloria telae." the ground truth keywords are very closely related to what we find using our approach. The term "information retrieval" is common in both. In the next example - "Visual encoding with Jittering eyes"- the topic "motion analysis" and "computer vision" are very closely related with the term human-computer interaction. Similarly, "bag-of-words model" is fre-quently used in "information retrieval" and "semantic memory" uses "World Wide Web". In the third example- "BuzzRank.. and the trend is your friend"- the topics "web community" and "e-social science" are highly relevant for "web graph" and "web dynamics".

*8.6 Results and discussion for keyword abstraction with de-noising*

This section is sub-categorized into two sections. The first section discusses the quan-titative evaluation of the proposed work. In the next section, we qualitatively discuss our results for some of the test documents.

Table 4: Table showing Jaccard Index measure for the proposed approach (varying k in context expansion) and the full content baseline.

| Clustering Algorithm | k=10 | k=20 | k=30 | k=40 | k=50 | Full Text* |
|---|---|---|---|---|---|---|
| Unpruned | 0.054 | 0.059 | 0.052 | 0.058 | 0.052 | 0.044 |
| Single | 0.054 | 0.057 | 0.050 | 0.057 | 0.056 | 0.040 |
| Complete | 0.058 | 0.055 | 0.043 | 0.047 | 0.052 | 0.034 |
| Average | 0.052 | 0.059 | 0.052 | 0.059 | 0.054 | 0.034 |

Table 5: Table showing results for a few of the sample documents. This table shows that several of the topics in the second column (our approach) are very closely related to the keywords in the ground truth (column 3).

| Document titles | Our approach | Ground truth |
|---|---|---|
| iTag: A Personalized Blog Tagger | web search,semantic technologies,semantic metadata,tag,meta data,computational linguistics, social bookmarking,data management | Tagging, Blogs, Machine Learning |
| Advances in Phonetic Word Spotting | speech recognition,language,linguistics,information retrieval,mobile phones,phoneme,speech processing, natural language processing,consonant,handwriting recognition,neural network | Speech recognition, synthesis Text analysis, Information Search and Retrieval |
| Mining the peanut gallery: opinion extraction and semantic classification of product reviews | linguistics,supervised learning,book review, unsupervised learning,review,parsing, sentiment analysis,machine learning | Opinion mining, document classification |
| Swoogle: A Search and Meta-data Engine for the Semantic Web | world wide web,search engine,web search engine, internet,social network, semantic search engine, search tools,semantic web,social networks,search engine optimization,ontology,web 2.0,semantics | Semantic Web, Search, Meta-data,Rank,Crawler |
| Factorizing Personalized Markov Chains for Next-Basket Recommendation | cold start,matrix, recommender systems, collective intelligence,markov chain, collaborative filtering, markov decision process | Basket Recommendation, Markov Chain, Matrix Factorization |

*8.6.1 Quantitative evaluation*

In order to compare the quality of keywords generated by both the approaches, we evaluate the results of the proposed approach and the baseline approach using the ground truth keywords for the test documents. The results of this experiment are shown in table 4. The first five columns correspond to the expanded context extracted using k as 10, 20, 30, 40 and 50. The Jaccard index of the baseline(Fulltext) with the ground truth is 0.044 whereas the Jaccard index for all the expanded context (proposed approach) over all values of k is greater than 0.50. The highest Jaccard index is 0.059 at k=20.

When we use the single hierarchical clustering algorithm for de-noising, the Jac-card index is only reduced to 0.040 for Fulltext baseline. The Jaccard index for the expanded context with k=20,40 is 0.057 which is clearly higher to the baseline re-sults. Similarly, for the complete hierarchical clustering based de-noising, the Jaccard index is 0.058 for k=10 whereas it is only 0.034 for the full text baseline. The same scenario is found for average hierarchical clustering based de-noising. The Jaccard index is 0.059 for k=20,40 while it is only 0.034 for Fulltext baseline.

The experiment described above shows a quantitative approach for comparing the quality of resultant keywords from the proposed and the baseline approaches.

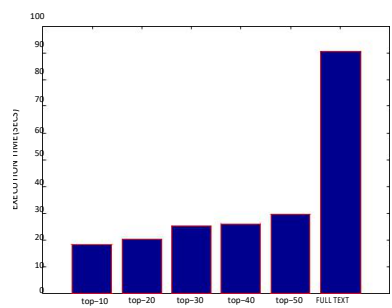

Fig. 9: Figure showing execution time comparison for the tag generation step using the expanded context (varying k) vs the full text for 50 documents.

The results shown above surprisingly favor the keywords generated by the proposed approach. The baseline approach uses the full text of the document in order to gener-ate keywords. An explanation for the observed results can be attributed to the fact that context derived from the web contains a wide spectrum of terms useful for generat-ing generalized tags for the document. While on the other hand, the full text approach uses only the terms local to the specific document which might not be diverse enough to generate generalized keywords.

One of the challenges described about using the full text approach is the issue of time consumption for reading the full text in case the document is large. In figure 9, we show the results of an experiment conducted to compare the execution time of the keyword generation step for the proposed and the baseline approaches. The x-axis in the figure shows the expanded context (using different values of k) and the baseline (Full text). The y-axis corresponds to the total execution time (in seconds) for 50 documents. As shown in the figure, the execution time for the baseline is approximately 90 seconds for 50 documents whereas the maximum execution time is only 30 seconds for the expanded context where k=50. As shown earlier, that the quality of keywords generated using expanded context with k=10 or k=20 is as good as higher values of k. This implies that good quality keywrods for a document can be generated 4.5 times faster using the proposed approach than using the full text of the document. This shows the effectiveness of the proposed approach to be useful in real time systems.

*8.6.2 Qualitative evaluation*

In this section, we discuss the results of the proposed approach by qualitatively an-alyzing the results of the proposed algorithm. The last section highlighted the per-formance of the proposed algorithm and quantitatively compared the results with the baseline using the Jaccard index. However, using a quantitative measure like Jaccard fails to account for the subjective accuracy of the tags other than those which do not match the ground truth exactly. Here we analyze the results subjectively.

Table 5 shows the keywords predicted by the proposed approach and the ground truth keywords for a few sample documents from the test dataset. For the first doc-ument in the table ('iTag: A Personalized Bog Tagger'), the keywords (our ground truth) assigned by the used contains terms like 'tagging', 'blogs' and 'Machine learn-ing'. Although there is no exact match between the proposed keywords and the ground truth keywords yet the relevance of the proposed keywords is striking. Key-words such as 'semantic meta data','social bookmarking', 'tag','computational lin-guistics' are similar others in this list are clearly good tags for this document. Another example is shown in the next row. The ground truth keyword 'speech recognition' ex-actly match the keyword in the proposed list. However, most of the other keywords in the list of proposed keywords are quite relevant. For example, tags such as 'lin-guistics','natural language processing' are closely related to this document. A few tags such as 'mobile phones','consonant,'hand writing recognition' may not be di-rectly related. The third example shown in this table also confirms the effectiveness of the proposed approach. The ground truth consists of only two tag: 'opinion min-ing' and 'document classification' while the proposed tag list consists several relevant tags though there is no exact match.

The last two examples shown in this table demonstrate the effectiveness of the approach to expand the keywords. The fourth example is originally tagged with key-words like 'semantic web','search','meta-data', 'rank' and 'crawler'. But the pro-posed list consists of highly relevant keywords like 'ontology', 'search optimization' which capture even the technique used in the particular research document. Similarly, for the last example the non-overlapping tags are relevant for annotating the research document.

## 9 Experimental analysis of the keyword extraction approach

The experimental setup is divided into two subsections. In the first part, we describe the preliminary analysis of the datasets used and explain the motivation for using the proposed approach. In the second part, we discuss the experimental design for evaluating the performance of the proposed approach. Before going into the details of the two sections, we first discuss about the datasets used in this work.

### 9.1 Dataset description
There are two datasets used for this approach. The datasets consist of research doc-uments from SIGKDD conference series[9], top-tier Data Mining conference in the computer science discipline. Each of the datasets consists of information about three attributes of the documents, namely, the title of the document, the abstract/summary text and the author assigned keywords. The information was obtained from the ACM digital library. The various statistics about the two datasets are given in table 6. Since the keyword vocabulary changes with time, we have selected datasets from different years to evaluate the performance of the proposed approach over time.

---

[9] http://www.sigkdd.org/

Table 6: Table summarizing the datasets used.

| Dataset id | Year | Count of documents |
|---|---|---|
| dataset$_1$ | 2008 | 118 |
| dataset$_2$ | 2012 | 112 |

Table 7: Results of preliminary analysis of the datasets.

| Dataset id | Total keywords | Keywords in titles | Keywords in abstract |
|---|---|---|---|
| dataset$_1$ | 438 | 3 | 208 (47.5%) |
| dataset$_2$ | 579 | 1 | 261 (45.0%) |

## 9.2 Preliminary analysis of the data

In this experiment, we performed a preliminary analysis of the above described datasets. The aim of the experiment is to derive statistics about the keywords used in the re-search documents. Table 7 summarizes the statistics collected from this experiment. As shown in this table, the dataset$_1$ consists of 438 keywords and 47.5% of these keywords appear in the abstract/ summary text of the document. However, the number of total keywords in dataset$_2$ are higher than in dataset$_1$. But even in this dataset, we find that 45% of the keywords appeared in the abstracts of the document. It should also be noted that negligible number of keywords appear in the title of the documents. From this analysis, we find that any keyword extraction algorithm can extract key-words from the abstract of the documents with lesser than 50% recall. So, in order to improve the recall of any algorithm, we need extra content information in addi-tion to the text in the abstract. Using full content of the document is one approach, however it has challenges such as (1) the computational time; and (2) risk of adding noise keywords[Lamb2008]. In such a scenario, the proposed approach provides a computationally efficient and superior solution (as shown in the following sections).

## 9.3 Experimental design parameters

In this section, we discuss the experimental parameters like the ground truth, the baselines, the evaluation metrics and various experiments to evaluate the performance of the proposed approach.

### 9.3.1 Ground truth

The ground truth is prepared from the author assigned keywords for the research doc-uments. We assume that the author assigned keywords are the best description of the summary of his document. In order to improve the comparison with the predicted keywords, the ground truth keywords are reduced to their root form using the stem-ming techniques. We used nltk (natural language toolkit) in python to stem the words to their root forms.

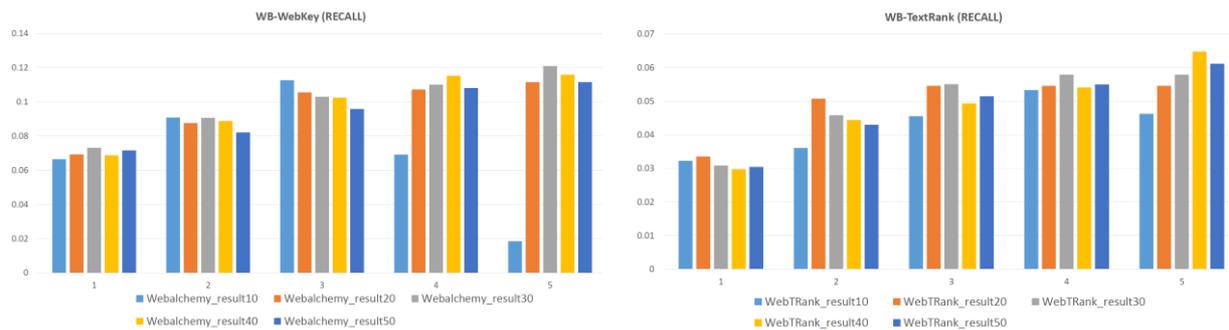

Fig. 10: Quantitative evaluation of n in the top-n titles used to create web-context. The plots show evaluation for (a) Alchemy and (b) TextRank techniques

### 9.3.2 Baselines

Since, the proposed approach creates extra content to improve keyword extraction, the approach can be used as a pre-step to any keyword extraction algorithm. Thus, the baselines in our experiments is the keyword extraction from the local text content of the document.

### 9.3.3 Evaluation metrics

We use precision and recall metrics to compare the performance of the proposed approach with the baselines. The precision and recall values are averaged over the number of documents in the datasets.

The experiments conducted in this work are aimed at bringing out three main understandings in relation to the performance of the proposed approach. We briefly describe the aim of the experiments in this section. The detailed analysis of the results in presented in the next section.

**Experiment 1** The aim of the experiment is to quantitatively evaluate the influence of n in the top n titles used in creating the web context used for a document.

**Experiment 2** The aim of the experiment is to quantitatively compare the performance of the three approaches for keyword extraction (Rake, TextRank and Alchemy).

**Experiment 3** The aim of the experiment is to quantitatively evaluate the impact of the web-context in improving the performance of keyword extraction techniques. The impact is evaluated by comparing with baselines.

### 9.4 Results and discussion

Figure 10(a) (b) shows the variation of recall@k for different values of n. As shown in this figure, the recall@k for k={5, 10, 15, 20, 25} is most higher when n=20,30. For other values of n, the recall is not consistently higher. It varies drastically when n=10. With n=50 also, the recall@k is not consistently high. Similar trends are observed in both the plots. From this analysis, we observe the variation of recall on the values of n selected to create the web-context. The analysis helps in determining the stable values of n that can be used for creating the web-context for optimal performance. For the experiments hence, we use n=20,30 for creating the web-context.

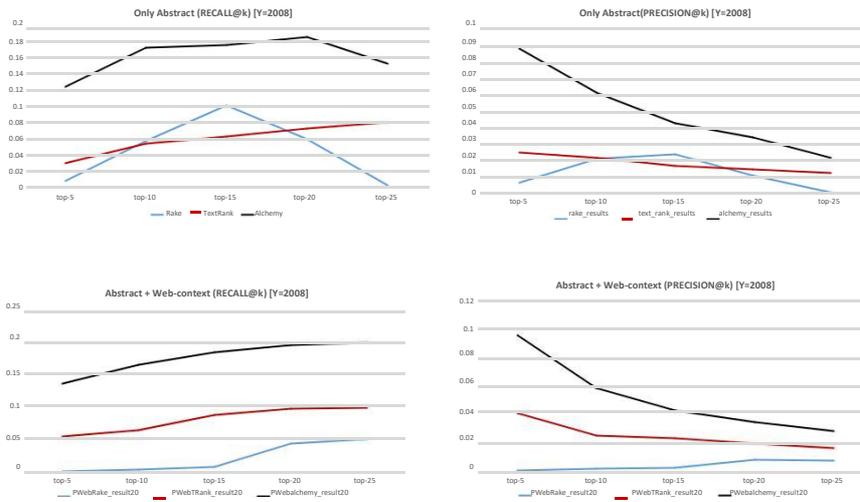

Fig. 11: Quantitative comparison of Rake, TextRank and Alchemy keyword extraction techniques for dataset$_1$. Plots (a, c) show the comparison via the recall@k metric. Plots (b, d) show the comparison via the precision@k metric. The value of k is in the range {5, 10, 15, 20, 25}. The results are shown for different document content (only abstract, only web-context and abstract+web-context). These figure show that the Alchemy keyword extraction technique (shown in black dash-dot) performs the best on both the metrics in all the cases.

Figure 11 (a-f) shows plots for quantitative comparison of Rake, TextRank and Alchemy keyword extraction approaches. Plots (a, c) show the comparison via the recall@k metric. Plots (b, d) show the comparison via the precision@k metric. The value of k is in the range f5, 10, 15, 20, 25g. The results are shown for different document content (only abstract and abstract+web-context). These figure show that the Alchemy keyword extraction technique (shown in black dash-dot) performs the best on both the metrics in all the cases.

As shown in the figure, the performance of the Alchemy technique is significantly higher than the Rake and TextRank techniques. These results shown in the figure are for dataset$_1$. However, the results for all the test datasets were consistent with these findings. This experiments reveals that the world knowledge based approach i.e. Alchemy technique is the best approach for keyword extraction. These results clearly demonstrate the advantage of incorporating information from the world knowledge corpus like Wikipedia to improve keyword extraction. Comparatively, the TextRank approach is clearly performing better than Rake in abstract+ Web context based key-word extraction. Therefore, in further analysis we will consider only TextRank and Alchemy techniques to monitor the impact of using web-context in addition to the summary text of the documents.

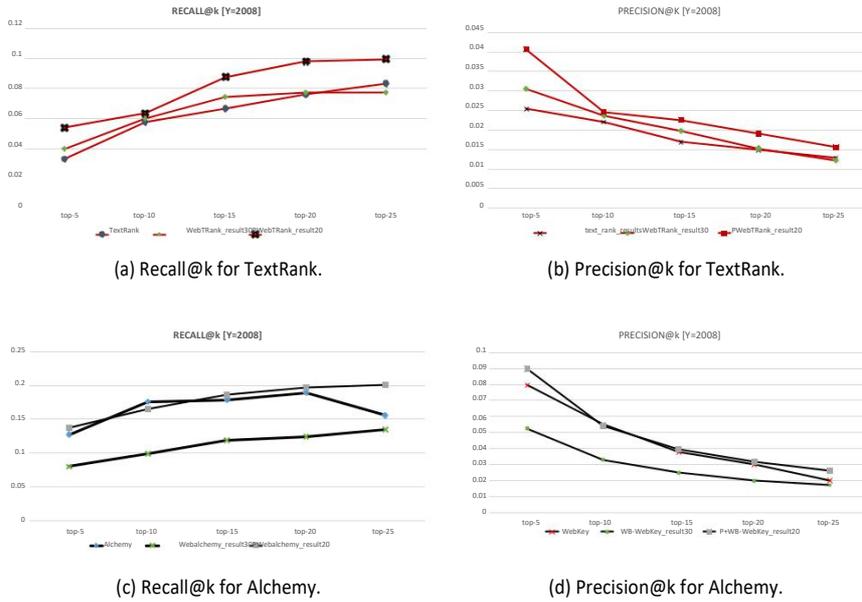

(a) Recall@k for TextRank.                      (b) Precision@k for TextRank.

(c) Recall@k for Alchemy.                      (d) Precision@k for Alchemy.

Fig. 12: Quantitative comparison of adding web-context to the local content for dataset$_1$ (2008). The figure shows the comparison using recall@k and precision@k metrics for TextRank and Alchemy techniques.

Figures 12 and 13 show the precision and recall results for comparing the performance of the proposed approach with the baselines for all the keywords extraction techniques for dataset$_1$ (KDD 08) and dataset$_2$ (KDD 12).

As shown in figure 12, the analysis is done to quantitatively evaluate the impact of adding web context to the summary text to enhance various keyword extraction approaches. In the plots (a) and (b), we compare the performance of the TextRank algorithm by varying the content used for keyword extraction. As shown in these plots, the proposed approach to add web context to abstract (Abstract + web-context) gives a 57% improvement in recall@5 and 60% improvement in precision@5. The proposed approach is shown with solid square marked red curve. The baseline is shown in short dashed red curve. From the analysis in the previous experiment, we know that the performance of the TextRank is better than the Rake algorithm for the datasets used in this work. Therefore, the improvements obtained using the proposed approach are significant.

In plots (c) and (d), we compare the performance of the Alchemy technique by varying the content used for keyword extraction. The solid and square marked black curve denotes the proposed approach and the short dashed black curve denote the abstract only baseline. As shown in the figure, the web-context improves the

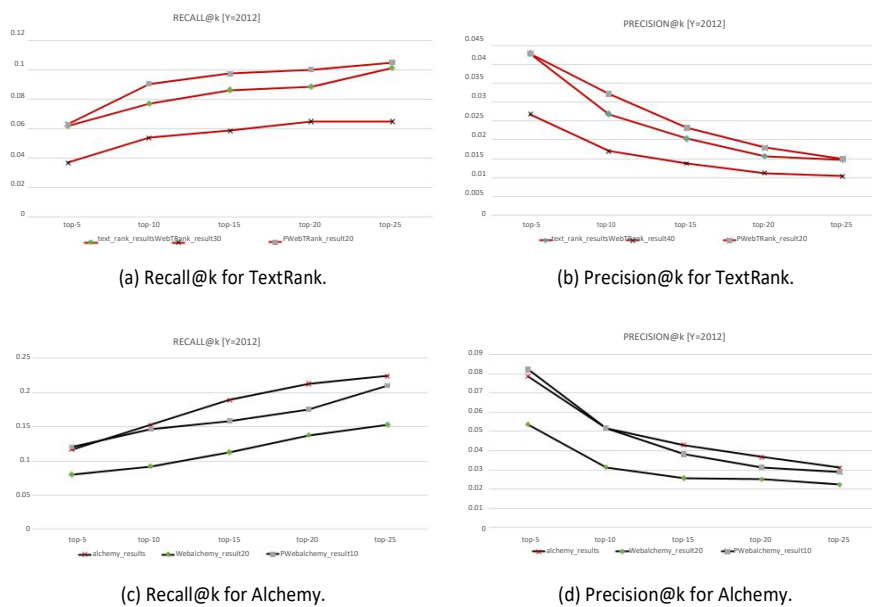

(a) Recall@k for TextRank.    (b) Precision@k for TextRank.

(c) Recall@k for Alchemy.    (d) Precision@k for Alchemy.

Fig. 13: Quantitative comparison of adding web-context to the local content for dataset$_2$ (2012). The figure shows the comparison using recall@k and precision@k metrics for TextRank and Alchemy techniques.

recall@5 by 11.3% and recall@25 by 33%. The precision@5 is also improved by 12.5% by using the web-context in addition to the text in the abstract of the documents.

These results on dataset$_1$ show interesting results demonstrating the significance of adding web-context to improve keyword extraction. However, we validate the per-formance of the proposed approach for dataset$_2$ as well. Figure 13 shows the results for dataset$_2$. In plots (a) and (b), we compare the performance of the TextRank al-gorithm by varying the content. The solid red and square marked curve denotes the performance of the proposed approach while the baseline (abstract only) approach is denoted by short red dashed and diamond marked curve. As shown in these plots, the recall@10 is boosted by approximately 17% and precision@10 is boosted by approx. 10%. These results demonstrate the significance of using the web-context to improve keyword extraction using the TextRank algorithm.

In plots (c) and (d), the performance of the Alchemy technique is compared by varying the content information. The performance of the proposed approach is demonstrated by a solid black and square marked curve while the baseline (ab-stract only content) is denoted by short black dashed and cross marked curve. The recall@5 shows an improvement of only 3% and the precision@5 improves by 4.6% over the baseline. The recall and precision values at k=10 are comparable for both the baseline and the proposed approach.

## 10 Conclusions

In summary, there are three main conclusions in this work. Firstly, we showed that the proposed keyword selection approach using only the title information of the doc-ument performs comparable with the baseline approach using the full text content of the document. The comparison was done using recall@k metric. Secondly, the fully automated approach for keyword assignment performs better than the baseline ap-proach in terms of the Jaccard index comparison. The proposed approach is atleast 3 times faster than the baseline approach using the full text content of the document to assign relevant keywords. There are several areas in this work to extend in the future. On of the areas of improvement in the current work is the de-noising algorithm which uses hierarchical clustering to pruning. However, hierarchical clustering has its lim-itations and it is worth to explore other algorithms such as density based clustering and some other novel anomaly detection algorithms. We would also test the proposed approach for other document corpus like news, patents etc.

Finally, we proposed a novel approach to improve the performance of the key-word extraction techniques. We performed several experiments to evaluate the im-provement in the performance of popular keyword extraction techniques by append-ing the web-context to the abstract of the document. By adding the web-context to the abstract of the document, we find that recall@5 and precision@5 are boosted by (approx.)60% for the TextRank algorithm. For the Alchemy algorithm, the proposed approach enhances the recall@5 by 11.3% and precision@5 by 12.5%. The perfor-mance was tested on different datasets.

## References

AlchemyAPI2013. AlchemyAPI. 2013. Text analysis by alchemyapi. http://www.alchemyapi.com.

Blei et al.2003. Blei, David M, Andrew Y Ng, and Michael I Jordan. 2003. Latent dirichlet allocation. the Journal of machine Learning research 3: 993–1022.

Browne1996. Browne, Glenda. 1996. Automatic indexing. Online Currents, the AusSI Newsletter 20 (6): 4–9.

Cilibrasi and Vitanyi2007. Cilibrasi, R. L., and P. M. B. Vitanyi. 2007. The google similarity distance.
IEEE Transactions on Knowledge and Data Engineering 19 (3): 370–383.

Clifton et al.2004. Clifton, Chris, Robert Cooley, and Jason Rennie. 2004. Topcat: data mining for topic identification in a text corpus. Knowledge and Data Engineering, IEEE Transactions on 16 (8): 949– 964.

DBLP. DBLP. Faceted dblp. http://dblp.l3s.de/browse.php?browse= mostPopularKeywords.

Dumais et al.1995. Dumais, Susan T, G Furnas, T Landauer, S Deerwester, S Deerwester, et al.. 1995. Latent semantic indexing. In Proceedings of the text retrieval conference.

Ertoz̈ et al.2003. Ertoz,̈ Levent, Michael Steinbach, and Vipin Kumar. 2003. Finding topics in collections of documents: A shared nearest neighbor approach. Clustering and Information Retrieval 11: 83–103.

Frank et al.1999. Frank, Eibe, Gordon W Paynter, Ian H Witten, Carl Gutwin, and Craig G Nevill-Manning. 1999. Domain-specific keyphrase extraction.

Gabrilovich and Markovitch2006. Gabrilovich, Evgeniy, and Shaul Markovitch. 2006. Overcoming the brittleness bottleneck using wikipedia: Enhancing text categorization with encyclopedic knowledge. In Aaai, Vol. 6, 1301–1306.

Hasan and Ng2010. Hasan, Kazi Saidul, and Vincent Ng. 2010. Conundrums in unsupervised keyphrase extraction: making sense of the state-of-the-art. In Proceedings of the 23rd international conference on computational linguistics: Posters, 365–373. Association for Computational Linguistics. Association for Computational Linguistics.

Hassan et al.2012. Hassan, Mostafa M, Fakhri Karray, and Mohamed S Kamel. 2012. Automatic docu-ment topic identification using wikipedia hierarchical ontology. In Information science, signal pro-cessing and their applications (isspa), 2012 11th international conference on, 237–242. IEEE. IEEE.

Hoffart et al.2013. Hoffart, Johannes, Fabian M Suchanek, Klaus Berberich, and Gerhard Weikum. 2013. Yago2: a spatially and temporally enhanced knowledge base from wikipedia. Artificial Intelligence 194: 28–61.

Jain and Pareek2010. Jain, Sonal, and Jyoti Pareek. 2010. Automatic topic (s) identification from learning material: An ontological approach. In Computer engineering and applications (iccea), 2010 second international conference on, Vol. 2, 358–362. IEEE. IEEE.

Joachims1998. Joachims, Thorsten. 1998. Text categorization with support vector machines: Learning with many relevant features. Springer.

Khabsa et al.2014. Khabsa, M, CL Giles, and Ren Zhang. 2014. The number of scholarly documents on the public web. PLoS ONE 9 (5): 93949.

Lamb2008. Lamb, James A. 2008. What is wrong with full text searches. http://www.jalamb.f9. co.uk/Full_text_searches.html.

Lawrie et al.2001. Lawrie, Dawn, W Bruce Croft, and Arnold Rosenberg. 2001. Finding topic words for hierarchical summarization. In Proceedings of the 24th annual international acm sigir conference on research and development in information retrieval, 349–357. ACM. ACM.

Ley and Reuther2006. Ley, Michael, and Patrick Reuther. 2006. Maintaining an online bibliographical database: The problem of data quality. In Egc, 5–10.

Lin1995. Lin, Chin-Yew. 1995. Knowledge-based automatic topic identification. In Proceedings of the 33rd annual meeting on association for computational linguistics, 308–310. Association for Compu-tational Linguistics. Association for Computational Linguistics.

Mihalcea and Tarau2004. Mihalcea, Rada, and Paul Tarau. 2004. Textrank: Bringing order into texts. In Proceedings of emnlp, Vol. 4. Barcelona, Spain. Barcelona, Spain.

Moura and Rezende2007. Moura, Maria Fernanda, and Solange Oliveira Rezende. 2007. Choosing a hi-erarchical cluster labelling method for a specific domain document collection. New Trends in Artificial Intelligence.

Page et al.1999. Page, Lawrence, Sergey Brin, Rajeev Motwani, and Terry Winograd. 1999. The pagerank citation ranking: Bringing order to the web..

Rose et al.2010. Rose, Stuart, Dave Engel, Nick Cramer, and Wendy Cowley. 2010. Automatic keyword extraction from individual documents. Text Mining.

Sebastiani2002. Sebastiani, Fabrizio. 2002. Machine learning in automated text categorization. ACM com-puting surveys (CSUR) 34 (1): 1–47.

Singhal2014. Singhal, Ayush. 2014. Leveraging open source web resources to improve retrieval of low text content items. PhD diss, UNIVERSITY OF MINNESOTA.

Singhal and Srivastava. Singhal, Ayush, and Jaideep Srivastava. Semantic tagging for documents using short textinformation.

Singhal and Srivastava2014a. Singhal, Ayush, and Jaideep Srivastava. 2014a. Generating semantic anno-tations for research datasets. In Proceedings of the 4th international conference on web intelligence, mining and semantics (wims14), 30. ACM. ACM.

Singhal and Srivastava2014b. Singhal, Ayush, and Jaideep Srivastava. 2014b. Leveraging the web for au-tomating tag expansion for low-content items. In Information reuse and integration (iri), 2014 ieee 15th international conference on, 545–552. IEEE. IEEE.

Singhal et al.2013. Singhal, Ayush, Ravindra Kasturi, and Jaideep Srivastava. 2013. Automating docu-ment annotation using open source knowledge. In Proceedings of the 2013 ieee/wic/acm interna-tional joint conferences on web intelligence (wi) and intelligent agent technologies (iat)-volume 01, 199–204. IEEE Computer Society. IEEE Computer Society.

Tiun et al.2001. Tiun, Sabrina, Rosni Abdullah, and Tang Enya Kong. 2001. Automatic topic identifica-tion using ontology hierarchy. In Computational linguistics and intelligent text processing, 444–453. Springer.

Turney2000. Turney, Peter D. 2000. Learning algorithms for keyphrase extraction. Information Retrieval 2 (4): 303–336.

Witten et al.1999. Witten, Ian H, Gordon W Paynter, Eibe Frank, Carl Gutwin, and Craig G Nevill-Manning. 1999. Kea: Practical automatic keyphrase extraction. In Proceedings of the fourth acm conference on digital libraries, 254–255. ACM. ACM.

Wu et al.2008. Wu, Ho Chung, Robert Wing Pong Luk, Kam Fai Wong, and Kui Lam Kwok.
    2008. Inter-preting tf-idf term weights as making relevance decisions. ACM Transactions
    on Information Systems (TOIS) 26 (3): 13.
Zamir and Etzioni1998. Zamir, Oren, and Oren Etzioni. 1998. Web document clustering: A
    feasibility demonstration. In Proceedings of the 21st annual international acm sigir
    conference on research and development in information retrieval, 46–54. ACM. ACM.